\documentclass[reprint, amssymb, amsmath,apl,aip]{revtex4-1}

\usepackage{bm}%
\usepackage{graphicx}

\begin{document}

\title{High-fidelity adiabatic inversion of a $^{31}\mathrm{P}$ electron spin qubit in natural silicon}%

\author{Arne Laucht}
\email{a.laucht@unsw.edu.au}
\affiliation{Centre for Quantum Computation and Communication Technology, School of Electrical Engineering \& Telecommunications, University of New South Wales, Sydney NSW 2052, Australia}
\author{Rachpon Kalra}
\affiliation{Centre for Quantum Computation and Communication Technology, School of Electrical Engineering \& Telecommunications, University of New South Wales, Sydney NSW 2052, Australia}
\author{Juha T. Muhonen}
\affiliation{Centre for Quantum Computation and Communication Technology, School of Electrical Engineering \& Telecommunications, University of New South Wales, Sydney NSW 2052, Australia}
\author{Juan P. Dehollain}
\affiliation{Centre for Quantum Computation and Communication Technology, School of Electrical Engineering \& Telecommunications, University of New South Wales, Sydney NSW 2052, Australia}
\author{Fahd A. Mohiyaddin}
\affiliation{Centre for Quantum Computation and Communication Technology, School of Electrical Engineering \& Telecommunications, University of New South Wales, Sydney NSW 2052, Australia}
\author{Fay Hudson}
\affiliation{Centre for Quantum Computation and Communication Technology, School of Electrical Engineering \& Telecommunications, University of New South Wales, Sydney NSW 2052, Australia}
\author{Jeffrey C. McCallum}
\affiliation{Centre for Quantum Computation and Communication Technology, School of Physics, University of Melbourne,
Melbourne VIC 3010, Australia}
\author{David N. Jamieson}
\affiliation{Centre for Quantum Computation and Communication Technology, School of Physics, University of Melbourne,
Melbourne VIC 3010, Australia}
\author{Andrew S. Dzurak}
\affiliation{Centre for Quantum Computation and Communication Technology, School of Electrical Engineering \& Telecommunications, University of New South Wales, Sydney NSW 2052, Australia}
\author{Andrea Morello}
\affiliation{Centre for Quantum Computation and Communication Technology, School of Electrical Engineering \& Telecommunications, University of New South Wales, Sydney NSW 2052, Australia}

\date{\today}%

\begin{abstract}
The main limitation to the high-fidelity quantum control of spins in semiconductors is the presence of strongly fluctuating fields arising from the nuclear spin bath of the host material. We demonstrate here a substantial improvement in single-qubit gate fidelities for an electron spin qubit bound to a $^{31}$P atom in natural silicon, by applying adiabatic inversion instead of narrow-band pulses. We achieve an inversion fidelity of 97~\%, and we observe signatures in the spin resonance spectra and the spin coherence time that are consistent with the presence of an additional exchange-coupled donor. This work highlights the effectiveness of adiabatic inversion techniques for spin control in fluctuating environments.
\end{abstract}

\maketitle
Spin qubits in semiconductors now represent one of the most promising solid-state architectures for quantum computation~\cite{Loss1998,Taylor2005,Hollenberg2006}, following the demonstration of coherent control of one-~\cite{Koppens2006} and two-~\cite{Petta2005} electron spin states in GaAs quantum dots and, more recently, singlet-triplet qubits in Si/SiGe dots~\cite{Maune2012} and $^{31}$P donor electron~\cite{Pla2012} and nuclear~\cite{Pla2013} spins in silicon.

In any III-V semiconductor, as well as in natural Si, the fluctuating nuclear spin environment is the main factor limiting spin coherence times~\cite{Yao2006,Witzel2010}, and, importantly, the fidelity of quantum gate operations, with typical fidelities in the range of 55 - 75 \% \cite{Koppens2006,Pla2012}. This is insufficient for fault-tolerant qubit operations, which require fidelities in excess of 99\% even in the most optimistic schemes~\cite{Wang2011}. Group IV semiconductors such as Si and C possess spin-zero nuclear isotopes, which can be artificially enriched to create a nearly spin-free environment for spin qubits. Indeed $^{28}$Si has been termed a \textquotedblleft semiconductor vacuum\textquotedblright~\cite{Steger2012} for this reason. Ensemble spin resonance of $^{31}$P donors in isotopically pure $^{28}$Si has shown extraordinarily long coherence times, $T_{2e} \approx 10$~s for the electron~\cite{Tyryshkin2012} and $T_{2n} \approx 3$~hours for the nucleus~\cite{Saeedi2013}, and it is certainly an exciting prospect to adopt isotopically pure substrates for nanoscale qubit devices. However, the production of nuclear spin-zero environments in isotopically purified semiconductors other than silicon is relatively undeveloped~\cite{Li2013} or impossible because of the lack of suitable isotopes. Therefore, methods to maximize qubit control fidelities in the presence of a nuclear spin environment will remain important.

In this letter, we present how diabatic and adiabatic frequency sweeps can be utilized to rotate the spin of an electron bound to a single $^{31}\mathrm{P}$ donor with high-fidelity, in spite of the fluctuating nuclear spins from the 4.7\% $^{29}Si$ (spin $1/2$) in natural silicon. For an inhomogeneously broadened electron spin resonance (ESR) transition with a linewidth of $\sim12$~MHz, we demonstrate that the use of adiabatic inversion allows us to achieve an average electron spin inversion fidelity as high as $F_I=97\pm 2$~\%. This inversion fidelity is insensitive to fluctuations of the background nuclear field, enriching our toolbox with an important technique for control sequences where high-fidelity electron spin inversions are required.

The sample investigated is similar to the one described in Pla \textit{et al.}~\cite{Pla2012}, and we refer to that publication for details about device fabrication and methods. The gate layout has been slightly modified, but the operation of the single electron transistor (SET) used for charge detection and the scheme used for spin-readout of the donor electron \cite{Morello2010a} is the same. There is, however, an important difference in the ion implantation method. In the present work we implanted $\mathrm{P}_2^+$ molecular ions, instead of $\mathrm{P}^+$ single ions. Upon impacting the Si chip, the $\mathrm{P}_2^+$ molecules break apart, leaving two separate P atoms at a distance that depends on the implantation energy. We used a $20$~keV acceleration voltage, which yields an expected average inter-donor distance of order $25$~nm~\cite{Ziegler10}. Our device then allows us to readout a single electron associated with one implanted P donor atom for our experiments.

\begin{figure}[t!]
\includegraphics[width=1\columnwidth]{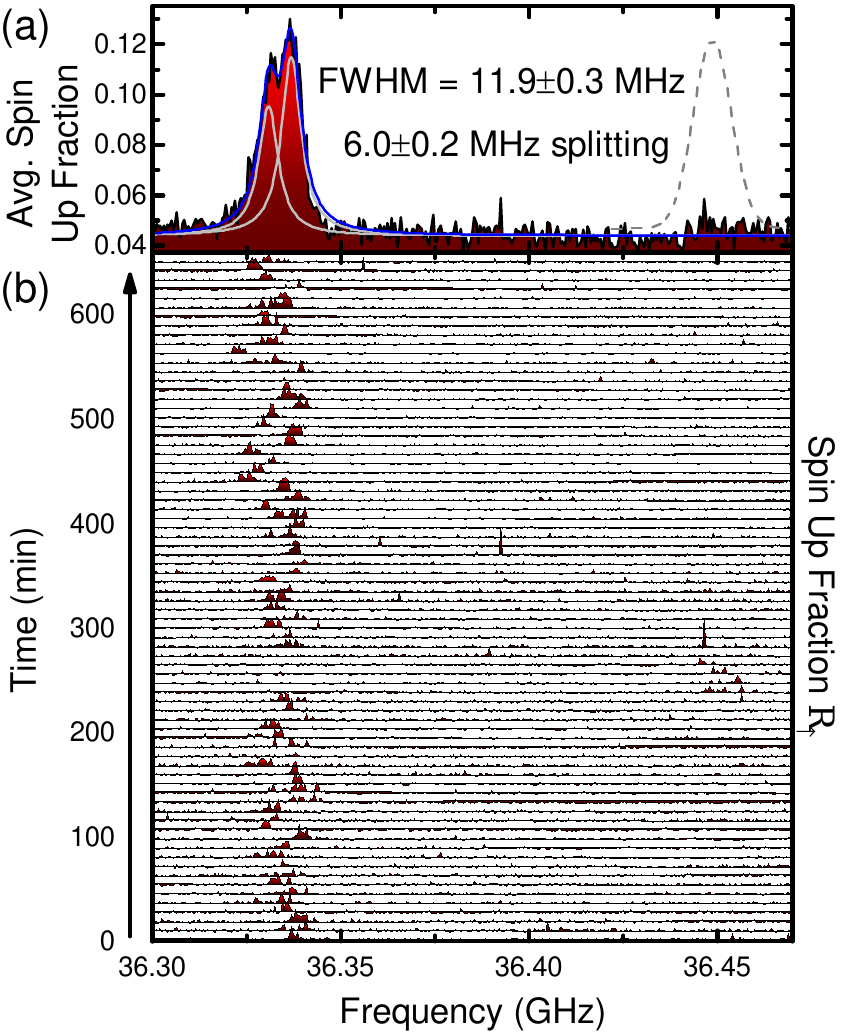}
\caption{\label{figure01} (a) Time-averaged electron spin resonance (ESR) spectrum of an electron bound to a $^{31}\mathrm{P}$ donor in natural silicon. The blue solid line is a fit to the data with two Gaussian peaks of equal width. The $^{31}$P nuclear spin was preferentially in the $|{\Downarrow\rangle}$ state during these measurements. Gray dotted line: expected ESR response for $|{\Uparrow\rangle}$ nuclear state, obtained from the hyperfine coupling $A_{\textsc{hf}}=114.4$~MHz determined from other measurements (data not shown). (b) Individual ESR spectra contributing to (a). These spectra were recorded over a time period of 660 minutes.}
\end{figure}

We start by discussing the measurement of the ESR spectra shown in Fig.~\ref{figure01}. These were obtained by monitoring the response of the electron spin to an applied microwave pulse. In the experimental sequence~\cite{Pla2012}, an electron with spin down $|{\downarrow\rangle}$ was loaded onto the donor. In a static magnetic field $B_0 = 1.3$~T the electron spin precesses with  a Larmor frequency $\nu_e = \gamma_e B_0 \pm A/2$ where $\gamma_e = 27.97$~GHz/T and $A$ is the hyperfine coupling to the $^{31}$P nucleus. A microwave pulse of power $P_{MW} = 2$~dBm at the source (the coaxial cable connecting the source to the device provides a further 30~dB attenuation) and duration $T_P=50$~$\mu$s was then applied to an adjacent broadband microwave antenna~\cite{Dehollain2012}. Since $T_P$ is much longer than the typical dephasing time $T_2^\star \approx 55$~ns for $^{31}\mathrm{P}$ in natural silicon~\cite{Pla2012}, the electron spin is left in a random orientation when the applied frequency is in resonance with the $|{\downarrow\rangle} \leftrightarrow |{\uparrow\rangle}$ ESR transition, or remains in the $|{\downarrow\rangle}$ state when off-resonance. In the last step, the orientation of the electron spin is read out in single-shot. The spectra in Fig.~\ref{figure01}(b) were recorded with $100$ repetitions at each frequency, before stepping to the next one. In this way, $75$ spectra were recorded over a time frame of $660$~minutes. The average of these 75 spectra is displayed in Fig.~\ref{figure01}(a), and appears as a broad distribution with a full-width at half-maximum  $\Delta\nu_{\textsc{fwhm}}=11.9\pm0.3$~MHz. The individual spectra in Fig.~\ref{figure01}(b) are much narrower than the average and show significant fluctuations in the position of the resonance, indicating a slow evolution of the $^{29}$Si nuclear field.

The averaged spectrum, however, is not a single resonance peak. It needs to be fitted with the sum (blue solid line) of two Gaussian peaks (gray solid lines) of equal width $\Delta\nu_{\textsc{fwhm}}=6.3\pm0.2$~MHz, separated by $\delta\nu=6.0\pm0.2$~MHz. This bimodal character suggests that the electron spin is coupled to a nearby two-level system that switches its state frequently over the timescale of the experiment. A splitting of $6$~MHz could either be caused by a hyperfine-coupled $^{29}\mathrm{Si}$ on the nearest-neighbor site \cite{Ivey1975}, or by another $^{31}\mathrm{P}$ donor, coupled to the donor under measurement by an exchange interaction $J = 14$~MHz, assuming that the two-electron system is initialized in the $|{\downarrow\downarrow\rangle}$ state~\cite{Kalra2013}. An exchange coupling of this magnitude corresponds to an inter-donor separation of $\sim 20$~nm~\cite{Wellard2003}, and is compatible with the expected inter-donor distance from 20~keV $\mathrm{P}_2^+$ molecular implantation. 

The donor under study is found predominantly in the nuclear $|{\Downarrow\rangle}$ state, but other ESR spectra in the nuclear $|{\Uparrow\rangle}$ state (data not shown) allowed us to determine the value of the hyperfine coupling  $A_{\textsc{hf}}=114.4$~MHz. This value is Stark-shifted from the bulk value of $117.53$~MHz~\cite{Feher59}, and is very close to the hyperfine splitting reported in Pla \textit{et al.}~\cite{Pla2012} and computed in Mohiyaddin \textit{et al.}~\cite{Mohiyaddin2013}.

The strong fluctuations and, therefore, the large broadening of the time-averaged ESR peak (cf. Fig.~\ref{figure01}) make it difficult to apply a microwave pulse in exact resonance with the instantaneous ESR frequency. High-fidelity single-qubit gate operations would require a Rabi frequency $\nu_1$ much larger than $\Delta\nu_{\textsc{fwhm}}$, which in the present case would translate into a strong rotating magnetic field $B_1 \gg \gamma_e \Delta\nu_{\textsc{fwhm}} \approx 0.23$~mT. Here we explore instead an easier and more reliable method based on adiabatic inversion.

The Landau-Zener theory~\cite{Zener32} applies to the time evolution of a two-level system described by a linearly-varying time-dependent Hamiltonian, where $h \nu_1$ ($2\nu_1$ corresponds to the Rabi frequency) couples the two levels. In our case the detuning $\Delta \nu$ between the source microwave frequency and the spin resonance is swept in time. The probability of a diabatic transition from one eigenstate to the other is given by:
\begin{equation}
\label{LZ}
P_D = \exp\left(-4\pi^2{\frac{\nu_1^2}{\left|\frac{\partial}{\partial t}\left(\Delta\nu\right)\right|}}\right).
\end{equation}
 When the rate of change of the energy difference (``sweep rate'', in frequency units) $\tfrac{\partial}{\partial t}\left(\Delta\nu\right)$ is low enough compared to the Rabi frequency $2\nu_1$, the system will adiabatically follow the instantaneous eigenstate. In the case of interest here, we consider an electron spin subject to a rotating magnetic field $B_1(t)$ at frequency $\nu_0(t) = \gamma_e B_0 - A/2 + \Delta \nu(t)$. Since the $^{31}$P nuclear spin remains in the ${|\Downarrow \rangle}$ state for several hours, we can treat the hyperfine field like a constant magnetic field shift, and describe the system in the $2 \times 2$ Hilbert space of the electron spin alone. In the reference frame rotating at frequency $\gamma_e B_0 - A/2$, the Hamiltonian of the system reads:
\begin{equation}
\label{hami}
H(t) = \frac{1}{2} \tfrac{\partial}{\partial t}\left(\Delta\nu\right) t \sigma_z + \nu_1 \sigma_x,
\end{equation}
where $\sigma_x$ and $\sigma_z$ are the spin Pauli matrices.
   
    An electron spin initialized in the $|{\downarrow\rangle}$ will be rotated to the $|{\uparrow\rangle}$ state once the frequency sweep $\Delta \nu \ll -\Delta\nu_{\textsc{fwhm}} \rightarrow \Delta \nu \gg \Delta\nu_{\textsc{fwhm}}$ is complete. For fast sweep rates, the spin state cannot perfectly follow the eigenstates, resulting in an incomplete inversion, i.e. a rotation of an angle $< \pi$. This method has been widely applied in nuclear magnetic resonance~\cite{Abragam1961} but less often in electron spin resonance, although recent progress in high-frequency electronics is making it more appealing~\cite{Doll2013}.

\begin{figure}[t!]
\includegraphics[width=1\columnwidth]{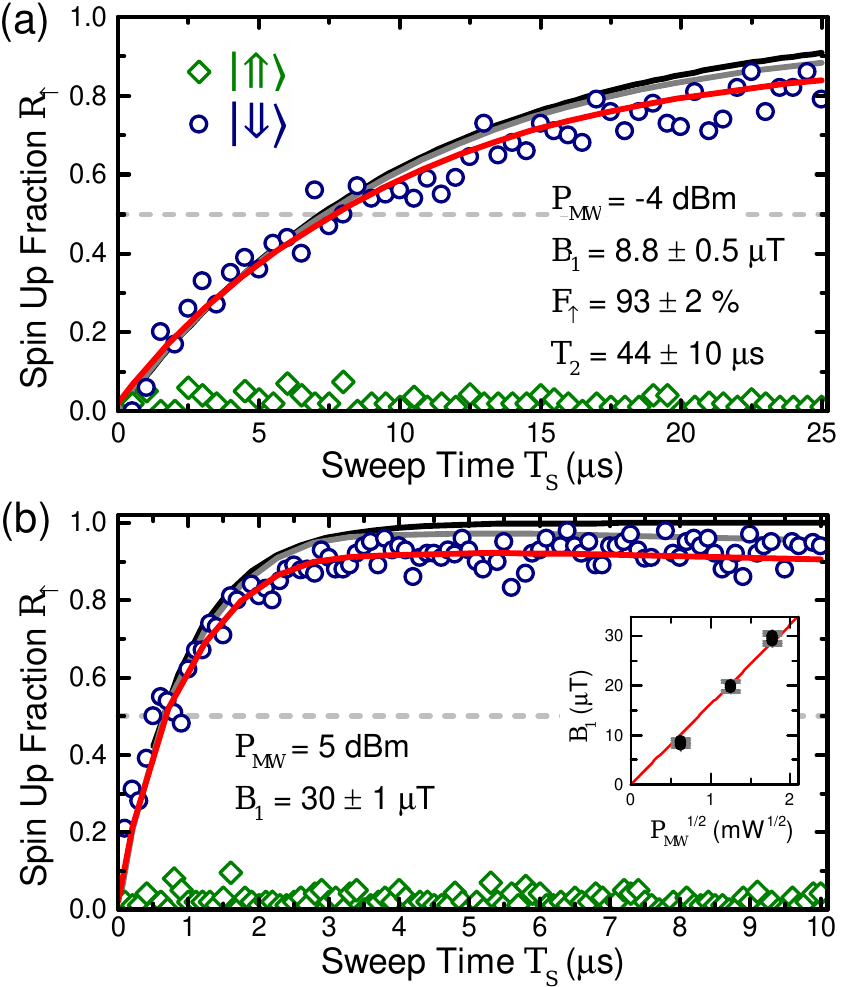}
\caption{\label{figure02} (a), (b) Electron spin-up fraction $R_\uparrow$ after a frequency sweep with duration $T_S$. The microwave frequency is swept over a range $\Delta\nu_{sweep}=25$~MHz, centered at the ESR frequencies for the $|{\Uparrow\rangle}$ (green diamonds) and $|{\Downarrow\rangle}$ (blue circles) nuclear states obtained from Fig.~\ref{figure01}. Black lines: response of an ideal system following the Landau-Zener formula~(\ref{LZ}). Gray lines: density matrix simulations of the diabatic sweep, accounting for finite $T_2$ spin coherence time. Red lines: including background counts $F_\uparrow P_{\uparrow I}$ and readout fidelity $F_{\uparrow}$. Dashed gray lines at $P_\uparrow = 0.5$ highlight the sweep time that would lead to a $\pi/2$ rotation. Data obtained with $-4$~dBm (a) and $5$~dBm (b) power of the microwave source, respectively. (inset) Extracted $B_1$ as a function of the square root of the applied microwave power. The red line line is a linear fit through the origin.}
\end{figure}

In Fig.~\ref{figure02} we present measurements of the electron spin-up fraction $R_\uparrow$ after loading an electron with spin down and performing a frequency sweep over a constant 25 MHz range, with variable sweep time $T_S$. The experiment in Fig.~\ref{figure02}(a) was conducted with a microwave power $P_{MW}=-4$~dBm at the source, while $P_{MW}=5$~dBm was used in Fig.~\ref{figure02}(b). For short sweep times, where $\tfrac{\partial}{\partial t}\left(\Delta\nu\right)$ is of the order of $4\pi^2 \nu_1^2$, the electron spin cannot adiabatically follow the instantaneous eigenstate. This regime can be used for controlled electron spin rotations of less than $\pi$. For example, a $\pi/2$ rotation is obtained for $T_S\sim7.5$~$\mu$s with $P_{MW}=-4$~dBm, and for $T_S\sim0.6$~$\mu$s with $P_{MW}=5$~dBm (dashed lines in Fig.~\ref{figure02}). For longer sweep times and larger microwave powers [see Fig.~\ref{figure02}(b)] the electron spin is fully inverted and the measured spin-up fraction $R_\uparrow$ saturates at a value close to unity, indicating high-fidelity spin inversion.

We model the experimental data using the density matrix formalism. The dephasing time $T_2$ of the electron spin is included in the master equation of the Lindblad form~\cite{Kok2010}
\begin{equation}
\label{lind}
\frac{d\rho}{dt}=-\frac{i}{\hbar}[H,\rho]+\mathcal{L}(\rho),
\end{equation}
where
\begin{eqnarray}
\label{liou}
\mathcal{L}(\rho) &= &\frac{1}{2 T_2} \left(2\sigma_z \rho \sigma_z - \sigma_z \sigma_z \rho - \rho \sigma_z \sigma_z \right)\nonumber \\ &= &\frac{2}{T_2} \begin{pmatrix}0 &-1\\-1 &0\end{pmatrix}.
\end{eqnarray}
We then use the equation of motion (\ref{lind}) to numerically compute the time evolution of an electron initialized in $|{\downarrow\rangle}$. For a meaningful comparison to the experiment, the model also incorporates: (i) the single-shot readout fidelity $F_\uparrow$ for the spin-up state; (ii) the background ``false counts'' rate $P_{\uparrow I} F_\uparrow$, where $P_{\uparrow I}$ is the probability that the electron tunnels out of the donor during the read-out phase, in the absence of ESR excitation~\cite{Pla2012}. While $F_\uparrow$ is a fitting parameter, $P_{\uparrow I} F_\uparrow$ is obtained from a measurement of $R_{\uparrow}$ while the $^{31}$P nuclear spin is in the ${|\Uparrow\rangle}$ state (green diamonds in Fig.~\ref{figure02}). In the present experiment, an electron temperature $T_{el} \approx 100$~mK allowed us to obtain a background count rate as low as 2.2$~$\%.

The results of our simulations are also plotted in Fig.~\ref{figure02}. The black solid line is simply the Landau-Zener formula (\ref{LZ}) which describes the response of an ideal system. The gray lines are density matrix simulations of the diabatic sweep, including only the $T_2$ time of the electron spin, while the red lines include the effect of background counts and readout fidelity. The best agreement with the experimental data is obtained by assuming a rotating magnetic field strength $B_1=8.8\pm0.5$~$\mu$T ($P_{MW}=-4$~dBm) and $B_1=30\pm1$~$\mu$T ($P_{MW}=5$~dBm), a readout fidelity  $F_\uparrow=93\pm2$~$\%$, and a decoherence time $T_2 = 44\pm10$~$\mu$s. Modeling a total of eight datasets for three different excitation powers (only two datasets shown) allows us to verify the $\sqrt{P_{MW}}$-dependence of $B_1$, confirmed by the good agreement of the extracted values [inset of Fig.~\ref{figure02}(b)] with a linear fit through the origin~\footnote{In contrast to the rotating $B_1$ field assumed in the simulations, the $B_1$ field in the experiment is linearly polarized. This results in the real $B_1$ field amplitude being a factor $2\times$ larger.}. For all simulations, $F_\uparrow$ and $T_2$ were global parameters, i.e. the same values were used for all the simulations. The decoherence time $T_2$ extracted from these data is significantly shorter than that for isolated single donors in natural silicon ($T_2 = 206$~$\mu$s in Ref.~[\onlinecite{Pla2012}]). Since exchange coupling creates an additional path for dephasing which can be very sensitive to electric field noise~\cite{Hu2006}, this short decoherence time further supports the possibility of having observed a weakly-coupled two-donor system, as first brought up by the bimodal shape of the ESR peak in Fig.~\ref{figure01}(a). However, the current status of this experiment does not allow us to conclusively exclude the possibility that these effects arise from a $^{29}$Si nucleus at the nearest-neighbor site.

From the gray line in Fig.~\ref{figure02}(b), which represents the result of the simulations without readout infidelity and background counts, we extract a maximum inversion fidelity of $F_I=97\pm2$~$\%$. This remarkable value is obtained for a moderate $B_1=30$~$\mu$T and a sweep time of $T_S \approx 6$~$\mu$s. Beyond this ($B_1$-dependent) optimal value of $T_S$, the inversion fidelity is deteriorated by the spin decoherence. $F_I$ would further increase with higher $B_1$ values, because the inversion could then be accomplished in a time $T_S \ll T_2$. A value of $F_I=97$~$\%$ represents a dramatic improvement when compared to the $61\pm 2$~$\%$ inversion fidelity~\footnote{The inversion fidelity was calculated as $F_I = 0.5 \textsc{cos}( (1-F_C) \pi )+0.5$, where $F_C=57\pm 2$~$\%$ is the angle control fidelity~\cite{Pla2012}.} obtained with resonant pulses in Ref.~[\onlinecite{Pla2012}], despite operating in the same $^{29}$Si nuclear spin environment.

In summary, we have presented high-fidelity adiabatic inversions of the spin of an electron bound to a $^{31}\mathrm{P}$ donor in natural silicon. Although the $^{29}\mathrm{Si}$ nuclear spins and, possibly, a second exchange coupled $^{31}$P donor, lead to an inhomogeneous broadening of the electron resonance frequency of $\sim12$~MHz, we are able to invert the electron spin with a fidelity of $F_I=97\pm2$~$\%$. This is made possible by the intrinsic robustness of this technique to the exact resonance frequency of the electron spin. Our result highlights the benefits of adiabatic inversion as the technique of choice for coherent control of spin qubits in environments that produce strong magnetic field fluctuations of nuclear origin, such as natural silicon and III-V semiconductors.

This research was funded by the Australian Research Council Centre of Excellence for Quantum Computation and Communication Technology (project number CE11E0096) and the US Army Research Office (W911NF-13-1-0024). We acknowledge support from the Australian National Fabrication Facility, and from the laboratory of Prof Robert Elliman at the Australian National University for the ion implantation facilities.

\bibliography{Papers}

\end{document}